\documentclass[5p, twocolumn]{elsarticle}
\usepackage[T1]{fontenc}
\usepackage{microtype}
\usepackage{amssymb}
\usepackage{amsmath}
\usepackage{multirow}
\usepackage{graphicx}
\usepackage{subcaption}
\usepackage[ruled,vlined,linesnumbered,algo2e]{algorithm2e}
\usepackage{url}

\usepackage{booktabs}

\journal{Integration, the VLSI Journal}

\begin{document}

\begin{frontmatter}

\title{Energy-Efficient FPGA Implementation of an End-to-End Cosine-Domain CNN Using High-Level Synthesis}

\author{Sana Fatima}
\author{Ab Al-Hadi Ab Rahman\corref{cor1}}
\ead{hadi@utm.my}
\author{Ibrahim Yousef Alshareef}

\author{Farhad Khodadadi}
\author{Nuzhat Khan}

\cortext[cor1]{Corresponding author}

\affiliation{
  organization={Universiti Teknologi Malaysia},
  department={Faculty of Electrical Engineering},
  addressline={VeCAD Research Lab},
  postcode={81310},
  state={Johor},
  country={Malaysia}
}

\begin{abstract}
The computational cost of conventional Convolutional Neural Networks (CNNs) limits their deployment in low-power devices for high-speed edge AI applications. This paper presents an end-to-end cosine-domain CNN (CD-CNN) architecture for energy-efficient FPGA inference using high-level synthesis (HLS). The proposed architecture performs intermediate feature extraction using real-valued DCT-domain representations, avoiding complex-valued arithmetic and intermediate inverse transforms between spectral layers. The design is implemented on a Xilinx Kintex UltraScale+ XCKU5P FPGA and evaluated against a fully spectral Fourier-domain CNN (FD-CNN) under consistent network and implementation constraints. Experiments on the MNIST and 94-class ASCII datasets show that the CD-CNN reduces normalized resource utilization and improves inference throughput on both datasets. Its energy efficiency reaches 112.91 frames/J for MNIST and 101.27 frames/J for ASCII, corresponding to improvements of $1.51\times$ and $2.76\times$, respectively, over the fully spectral FD-CNN architecture. Furthermore, compared to the previous GPU-based fully spectral CD-CNN implementation, the proposed end-to-end FPGA implementation improves energy efficiency by $12.96\times$. These results demonstrate the potential of real-valued spectral processing in the DCT domain for energy-efficient edge AI deployment.
\end{abstract}

\begin{keyword}
Discrete Cosine Domain\sep
Convolutional Neural Network\sep
Field Programmable Gate Arrays \sep
High-Level Synthesis \sep
Energy Efficiency
\end{keyword}

\end{frontmatter}

\section{Introduction}

The rapid advancement of artificial intelligence (AI) has led to the widespread integration of deep learning models across numerous edge applications. Driven by architectural advances, convolutional neural networks (CNNs) have become a common choice for image and signal processing tasks \cite{alshareef2025lightweight, chen2021review}. However, executing these computationally intensive networks on resource-constrained hardware introduces significant design challenges.

General-purpose computing platforms, including graphics processing units (GPUs), can incur substantial power overhead, while central processing units (CPUs) may not meet the throughput requirements of some CNN workloads. Although application-specific integrated circuits (ASICs) can provide high efficiency, their development cost and design time can be prohibitive. Field-programmable gate arrays (FPGAs) offer a flexible platform for CNN acceleration through parallel processing and application-specific hardware optimization.

Efficient edge acceleration therefore requires algorithms that operate within limited resource and energy budgets. Conventional CNNs rely on sliding-window convolution, which can impose substantial computational and data-movement costs. To reduce convolutional overhead, previous studies have proposed spectral CNNs (SpCNNs) that use element-wise multiplication in a transformed domain \cite{mathieu2014fast, rippel2015spectral}. Among spectral transforms, the Fast Fourier Transform (FFT) has been widely used in prior work \cite{ayat2019spectral, rizvi2021low, rizvi2023computation, abtahi2018accelerating}, partly because of its integration with deep-learning frameworks and optimized implementations \cite{he2020fft}. Previous Fourier-domain CNN (FD-CNN) FPGA implementations \cite{zhang2017frequency, zeng2017fast, zeng2018framework, abtahi2018accelerating, sun2018throughput, niu2019spec2, liu2020optimizing, liu2022design} have predominantly used register-transfer-level (RTL) design. RTL can provide fine-grained hardware control, but it requires substantial development effort. Therefore, this work uses high-level synthesis (HLS) to streamline hardware development while evaluating the resulting implementation at the lower level.

However, FFT-based models generally use complex-valued data, increasing arithmetic and storage requirements in constrained hardware \cite{niu2019spec2}. In contrast, the discrete cosine transform (DCT) uses real-valued coefficients and can therefore reduce the overhead associated with complex-valued processing \cite{xu2021dct}. A key challenge in SpCNNs is the repeated transformation between spatial and spectral domains, which can offset the benefits of spectral convolution. This work adopts a fully spectral DCT-domain CNN algorithm that integrates DCT-ReLU and Layer Adaptive Spectral Pooling (LASP), introduced by Alshareef \textit{et al.} \cite{alshareef2025real, alshareef2026relu}. This approach maintains intermediate feature representations in the DCT domain while retaining a lightweight network architecture.

Despite advances in DCT-domain CNN algorithms, their end-to-end realization on FPGA hardware remains an important implementation challenge. An effective hardware implementation must integrate input transformation, spectral feature extraction, activation, pooling, intermediate data storage, and classification while accounting for the resource and timing constraints of the target device. Furthermore, the hardware implications of real-valued DCT-domain processing have not been sufficiently established through a controlled comparison with fully spectral Fourier-domain implementations under consistent implementation conditions. This work addresses these gaps by implementing an end-to-end CD-CNN using HLS and evaluating its hardware characteristics against a corresponding FD-CNN on the same FPGA platform. To the best of our knowledge, this work presents the first end-to-end, fully spectral CD-CNN implementation on an FPGA using HLS.

The main contributions of this paper are as follows:
\begin{itemize}
    \item[(1)] An HLS-based harwdare architecture for fully spectral DCT-domain CNN inference, integrating input transformation, real-valued spectral convolution, DCT-ReLU, LASP, on-chip feature-map buffering, and fully connected classification without intermediate inverse transforms.
    \item[(2)] An evaluation of the proposed CD-CNN against a fully spectral FD-CNN under consistent network configurations and implementation conditions, quantifying the differences in FPGA resource utilization, inference throughput, power consumption, and energy efficiency.
    \item[(3)] An evaluation on MNIST and a 94-class ASCII dataset, demonstrating the classification performance and hardware efficiency of the proposed architecture across two classification tasks.
\end{itemize}

The remainder of this paper is organized as follows: Section 2 reviews related work. Section 3 describes the methodology. Section 4 presents the experimental results, Section 5 compares the proposed architecture with previous works, and Section 6 concludes the paper.

\section{Related Work}

Conventional CNNs use sliding-window filters to learn image features, which can impose high computational costs \cite{mathieu2014fast}. Spectral variants instead perform convolution through element-wise multiplication in a transformed domain. However, repeated transformations between spatial and spectral representations can reduce the computational benefits of spectral convolution \cite{rizvi2023computation}.

To address this issue, researchers have developed spectral-domain activation functions \cite{ayat2019spectral, rizvi2021low}, batch normalization \cite{ayat2019spectral}, and pooling operations \cite{rippel2015spectral}. Some architectures still apply an inverse FFT before classification, whereas fully spectral designs maintain intermediate representations in the Fourier domain \cite{fulop2022end}.

Despite their computational advantages, Fourier-domain hardware implementations can require substantial resources for complex-valued processing \cite{zhang2017frequency, zeng2017fast, zeng2018framework, abtahi2018accelerating, sun2018throughput, niu2019spec2}. Furthermore, these implementations accelerate convolution operations while executing input transformation, activation, pooling, and classification on the host CPU. Xu \textit{et al.} \cite{xu2021dct} investigated DCT-based convolution as an alternative to complex-valued Fourier-domain processing. Subsequent work by Alshareef \textit{et al.} \cite{alshareef2025real, alshareef2025optimizing, alshareef2026end, alshareef2026relu} developed CD-CNN algorithms that maintain real-valued feature extraction in the DCT domain. However, these advancements in DCT domain CNNs remain constrained to the algorithmic level. 

Previous SpCNN FPGA implementations have primarily relied on custom RTL, which requires expertise in hardware description languages and can increase design effort \cite{lahti2025high, TRETS2025HLS}. This work uses HLS to implement an end-to-end CD-CNN in C++, reducing the need for manually specified RTL while enabling post-implementation evaluation.

\section{Methodology}

\subsection{Overall Framework}

\begin{figure}[t]
    \centering
    \includegraphics[width=0.8\linewidth]{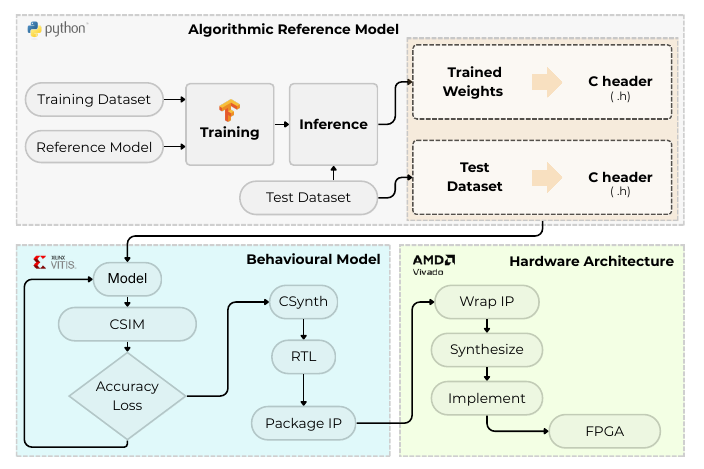}
    \caption{Hardware-Software Co-design framework.}
    \label{fig:overall_framework}
\end{figure}

This work presents an end-to-end FPGA implementation of a fully spectral CD-CNN and evaluates it against a fully spectral FD-CNN. As shown in Figure~\ref{fig:overall_framework}, the reference models are trained and validated in TensorFlow. The trained weights and representative test inputs are exported as C header files for the HLS implementations. Functional correctness is checked through C simulation, followed by C synthesis and RTL co-simulation. The generated RTL is integrated into the Vivado design flow for implementation, timing analysis, resource evaluation, and power estimation.

The architectures are compared in terms of classification accuracy, inference throughput, FPGA resource utilization, power consumption, and energy efficiency. Both architectures are derived from a common reference model and implemented using the same HLS design flow on the AMD Xilinx Kintex UltraScale+ XCKU5P FPGA. Network configuration, training procedure, datasets, numerical precision, and implementation constraints are controlled to support a fair comparison.

\subsection{Reference Model and Network Configuration}

\begin{table}[t]
\caption{Common reference configuration for FD-CNN and CD-CNN.}
\label{tab:reference_config}
\centering
\footnotesize
\begin{tabular}{ll}
\toprule
\textbf{Parameter} & \textbf{Configuration} \\
\midrule
Network depth & 3 convolutional layers \\
Kernel configuration & $[2,4,16]$ \\
Training framework & TensorFlow \\
Training epochs & 80 \\
Batch size & 32 \\
Optimizer & Adam \\
Learning rate & $1\times10^{-4}$ \\
Datasets & ASCII and MNIST \\
Precision & 32-bit floating-point \\
HLS framework & AMD Xilinx Vitis HLS 2025.2 \\
FPGA platform & AMD Xilinx Kintex UltraScale+ XCKU5P \\
\bottomrule
\end{tabular}
\end{table}

\begin{figure}[t]
    \centering
    \begin{subfigure}[t]{0.48\textwidth}
        \centering
        \includegraphics[width=\linewidth]{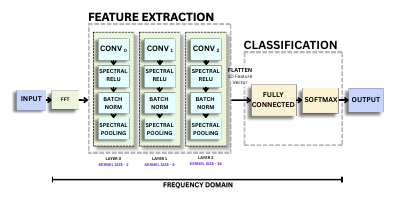}
        \caption{Fourier Domain CNN.}
        \label{fig:fft_reference}
    \end{subfigure}
    \hfill
    \begin{subfigure}[t]{0.48\textwidth}
        \centering
        \includegraphics[width=\linewidth]{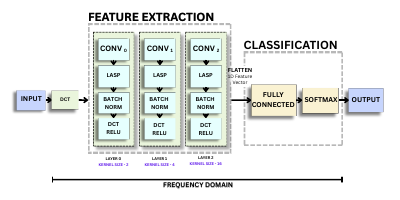}
        \caption{Cosine domain CNN.}
        \label{fig:dct_reference}
    \end{subfigure}
    \caption{Algorithmic Reference Models.}
    \label{fig:REFERENCE}
\end{figure}

The reference model in Figure~\ref{fig:REFERENCE} provides a common benchmark for the evaluated spectral CNNs. The network configuration is held constant across the FD-CNN and CD-CNN implementations, as summarized in Table~\ref{tab:reference_config}. The models use three convolutional layers with kernel configuration $K=[2,4,16]$. This lightweight configuration provides a common basis for evaluating the hardware implications of the two spectral representations.

The evaluated models use the 94-class ASCII dataset and the MNIST dataset. The ASCII dataset contains 23,848 samples across 94 character classes, including uppercase and lowercase letters, digits, and special characters \cite{alshareef2025lightweight}. MNIST contains 70,000 handwritten digit images across ten classes. The common reference model provides a controlled basis for evaluating the effects of the spectral domain on classification and hardware implementation.

\subsection{CD-CNN Algorithm}
Although the FD-CNN and CD-CNN share a common network configuration, their convolution, activation, and pooling operations are implemented using algorithms suited to their respective domains.

\subsubsection{Spectral Convolution}

\begin{figure*}[t]
    \centering
    \begin{subfigure}[t]{0.48\textwidth}
        \centering
        \includegraphics[width=\linewidth]{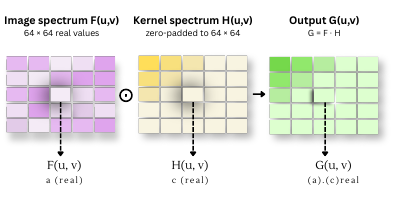}
        \caption{Discrete cosine-domain CNN.}
        \label{fig:dct_conv}
    \end{subfigure}
    \hfill
    \begin{subfigure}[t]{0.48\textwidth}
        \centering
        \includegraphics[width=\linewidth]{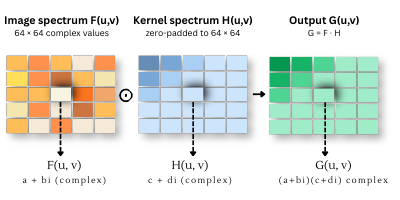}
        \caption{Fourier-domain CNN.}
        \label{fig:fft_conv}
    \end{subfigure}

    \caption{Spectral convolution operations.}
    \label{fig:conv_comparison}
\end{figure*}

The adopted SpCNN models are designed to avoid intermediate domain transformations. The proposed CD-CNN performs convolution in the DCT domain using real-valued coefficients, as shown in Figure \ref{fig:dct_conv}. Let \(X_D(u,v)\) and \(K_D(u,v)\) denote the DCT-domain representations of the input feature map and zero-padded convolution kernel, respectively. Following DCT-based spectral convolution formulations \cite{ju2022convolutional,ito2022convolution, alshareef2026end}, the operation is expressed as
\[
G_D(u,v)=X_D(u,v)\odot K_D(u,v),
\]
where \(\odot\) denotes element-wise multiplication. Each DCT-domain convolutional layer performs coefficient-wise multiplication between the input feature maps and learned spectral weights, followed by accumulation across input channels and bias addition. The resulting feature maps are processed by batch normalization and DCT-ReLU activation.

The FD-CNN uses Fourier-domain representations of input feature maps and convolution kernels, as shown in Figure \ref{fig:fft_conv}. Let $X_F(u,v)$ and $K_F(u,v)$ denote their Fourier-domain representations. Spectral convolution is computed using complex-valued element-wise multiplication:
\begin{equation}
    G_F(u,v)=X_F(u,v)\odot K_F(u,v),
    \label{eq:fft_spectral_conv}
\end{equation}
where $\odot$ denotes element-wise multiplication \cite{fulop2022end}. Unlike the FD-CNN, the CD-CNN uses real-valued coefficients and therefore does not require separate real and imaginary datapaths. This simplifies the arithmetic representation in CD-CNN, thereby reducing the intermediate storage requirements and computations during edge deployment.

\subsubsection{Spectral Pooling}

\begin{figure}[t]
    \centering
    \includegraphics[width=1\linewidth]{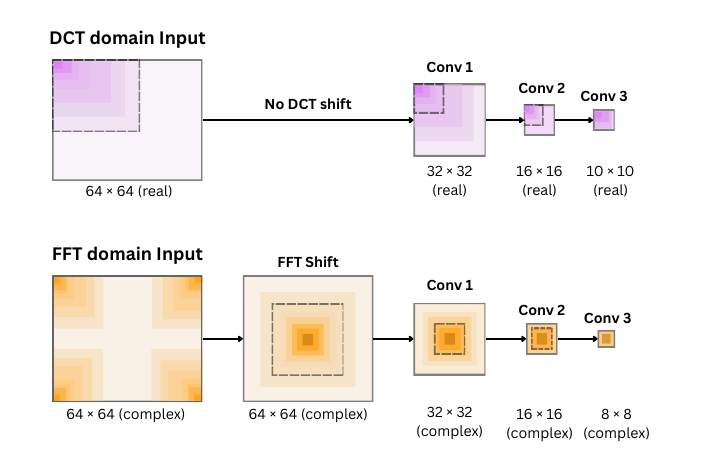}
    \caption{Spectral pooling for FD-CNN and CD-CNN architectures.}
    \label{fig:pooling}
\end{figure}

Figure~\ref{fig:pooling} presents the spectral pooling strategies used by the FD-CNN and CD-CNN models. Both architectures reduce the feature-map dimensions by retaining selected spectral coefficients, but the coefficient arrangements differ because Fourier and DCT representations have different frequency organizations.

In the CD-CNN, LASP retains the selected DCT coefficients without applying a Fourier-style frequency shift. The retained coefficients form a reduced representation for the next convolutional stage, while the discarded coefficients are excluded from subsequent computation. As shown in Figure~\ref{fig:pooling}, LASP reduces the CD-CNN feature-map dimensions by retaining 50\% for the first 2 layers and 65\% for the last layer, from $64\times64$ at the input to $32\times32$, $16\times16$, and $10\times10$ across the three convolutional stages. The FFT-domain architecture uses spectral pooling, reducing the feature map dimensions successively by half \cite{rippel2015spectral}, as illustrated in the same figure. The minor discrepancies in the output feature map dimensions between the two architectures were introduced to ensure a comparable parameter count.

By reducing the number of coefficients processed and stored in subsequent layers, spectral pooling reduces the computational workload and intermediate-buffer requirements. Combined with real-valued DCT coefficients, this design avoids the storage overhead associated with imaginary components. The feature maps remain in the spectral domain between layers, avoiding intermediate inverse transforms.

\subsection{CD-CNN Hardware Architecture}

The CD-CNN hardware architecture implements the DCT-domain operations described in the algorithm using real-valued arithmetic. Unlike the FD-CNN, which processes complex-valued coefficients, the CD-CNN uses real-valued DCT coefficients, avoiding separate real and imaginary datapaths.

\subsubsection{Spectral Convolution Datapath}

The input image is transformed using a two-dimensional DCT before entering the convolutional layers. As shown in Figure~\ref{fig:conv_comparison}, each layer multiplies input coefficients by learned spectral weights and accumulates the results across input channels. Batch normalization and DCT-ReLU are then applied. LASP retains the selected low-frequency coefficients between layers, reducing the number of coefficients processed by subsequent stages.

\subsubsection{On-Chip Dataflow and Control}

\begin{figure*}[t]
    \centering
    \includegraphics[width=\textwidth]{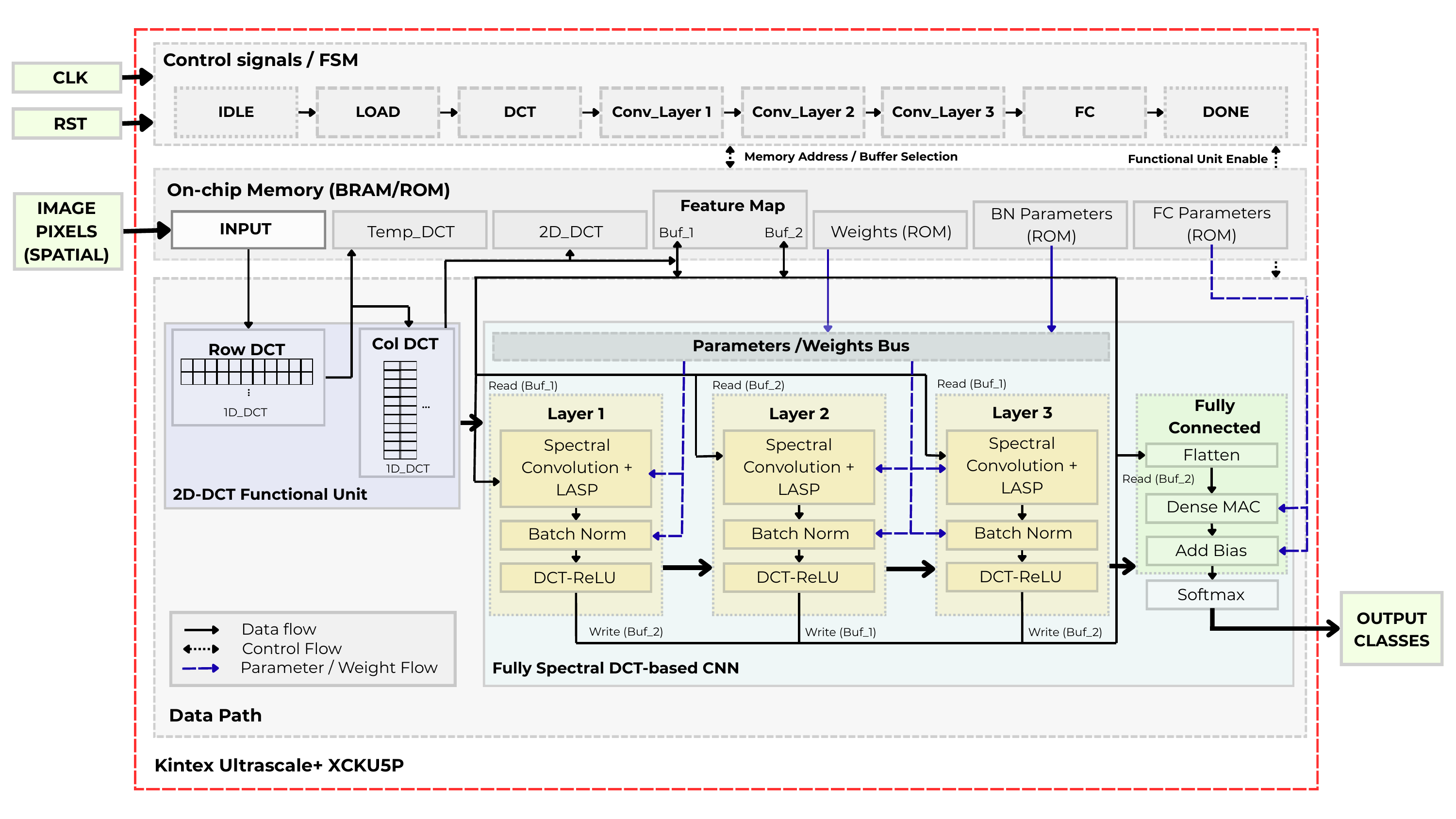}
    \caption{Hardware architecture of the proposed CD-CNN accelerator.}
    \label{fig:dct_hardware_architecture}
\end{figure*}

Figure~\ref{fig:dct_hardware_architecture} presents the architecture of the proposed accelerator on the Xilinx Kintex UltraScale+ XCKU5P FPGA. The architecture consists of a Finite State Machine (FSM)-based control unit, an on-chip memory subsystem, a two-dimensional DCT functional unit, and a spectral CNN datapath containing three convolutional stages and a fully connected classifier.

The control unit sequences execution through the IDLE, LOAD, DCT, convolutional-layer, FC, and DONE states. It coordinates memory addressing, buffer selection, and functional-unit activation. The on-chip memory subsystem stores input data, intermediate feature maps, spectral weights, batch-normalization parameters, and fully connected-layer parameters. Intermediate feature maps are maintained in on-chip buffers, while weights and layer parameters are accessed from ROM.

During inference, the spatial-domain input image is loaded and processed by the two-dimensional DCT unit. Row-wise and column-wise DCT operations produce the real-valued spectral representation required by the first convolutional stage. Each convolutional stage then performs spectral convolution fused with LASP, batch normalization, and DCT-ReLU. The same processing sequence is applied to the subsequent stages, with each stage operating on the reduced spectral representation produced by its predecessor.

Two buffers, Buf\_1 and Buf\_2, alternate between storing layer inputs and outputs. This ping-pong arrangement prevents output data from overwriting the current input feature map. After the final convolutional layer, the retained coefficients are flattened and passed to the fully connected classifier to generate the output class probabilities via softmax. Intermediate feature maps remain on-chip throughout inference.

Overall, the architecture performs domain transformation, real-valued DCT-domain convolution operation, DCT-domain spectral pooling and activation, using ping-pong buffering. These choices reduce the need for complex arithmetic, limit intermediate storage, and reduce data movement relative to the corresponding complex-valued spectral implementation.

\subsection{Evaluation Metrics}

The hardware implementations are evaluated using inference throughput, normalized resource utilization, power consumption, and energy efficiency.

\subsubsection{Throughput}

The throughput for single-frame inference is defined as

\begin{equation}
    \mathrm{Throughput}_{\mathrm{FPS}}=
    \frac{1}{\mathrm{Latency}_{s}}
    \label{eq:throughput}
\end{equation}
Frames per second (FPS) is used as an application-level measure of inference throughput.

\subsubsection{Normalized Resource Utilization}

To compare different FPGA resource types using a common estimate, resource consumption is expressed as a CLB-equivalent metric. If a CLB is represented by $N_{\mathrm{FF}}$ flip-flops and $N_{\mathrm{LUT}}$ lookup tables, and DSP and BRAM resources are converted using weighting factors $W_{\mathrm{DSP}}$ and $W_{\mathrm{BRAM}}$, respectively, the estimated resource metric is
\begin{equation}
\label{eq:clb_normalized}
N_{\mathrm{CLB}}^{\mathrm{eq}} =
\max\left(
\frac{\mathrm{LUT}}{N_{\mathrm{LUT}}},
\frac{\mathrm{FF}}{N_{\mathrm{FF}}}
\right)
+W_{\mathrm{DSP}}\mathrm{DSP}
+W_{\mathrm{BRAM}}\mathrm{BRAM}
\end{equation}

For the XCKU5P device, this work uses $N_{\mathrm{FF}}=16$, $N_{\mathrm{LUT}}=8$, and Vivado post-implementation layout view estimates $W_{\mathrm{DSP}} = W_{\mathrm{BRAM}} = 5$ CLB slices, based on the adopted normalization procedure and the relevant AMD device documentation \cite{amd_ug574}. The total device reference is $N_{\mathrm{CLB}}^{\mathrm{total}}=38{,}640$. The normalized resource utilization is calculated as
\begin{equation}
    Utilization_{\%}=
    \frac{N_{\mathrm{CLB}}^{\mathrm{eq}}}
    {N_{\mathrm{CLB}}^{\mathrm{total}}}\times100\%
    \label{eq:resource_util}
\end{equation}
This value is an estimated CLB-equivalent metric, not a direct count of physically occupied CLBs. The normalization assumptions should be interpreted accordingly.

\begin{figure*}[t]
    \centering
    \begin{subfigure}[t]{0.48\textwidth}
        \centering
        \includegraphics[width=\linewidth]{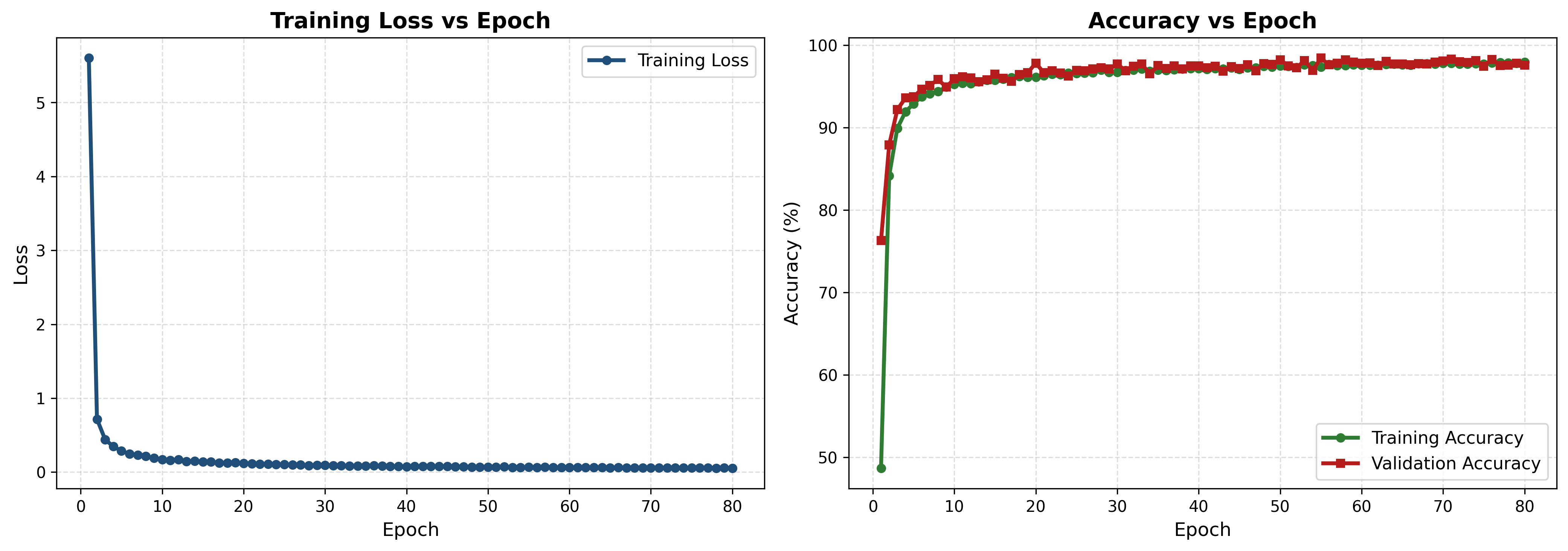}
        \caption{ASCII dataset.}
        \label{fig:fft_ascii}
    \end{subfigure}
    \hfill
    \begin{subfigure}[t]{0.48\textwidth}
        \centering
        \includegraphics[width=\linewidth]{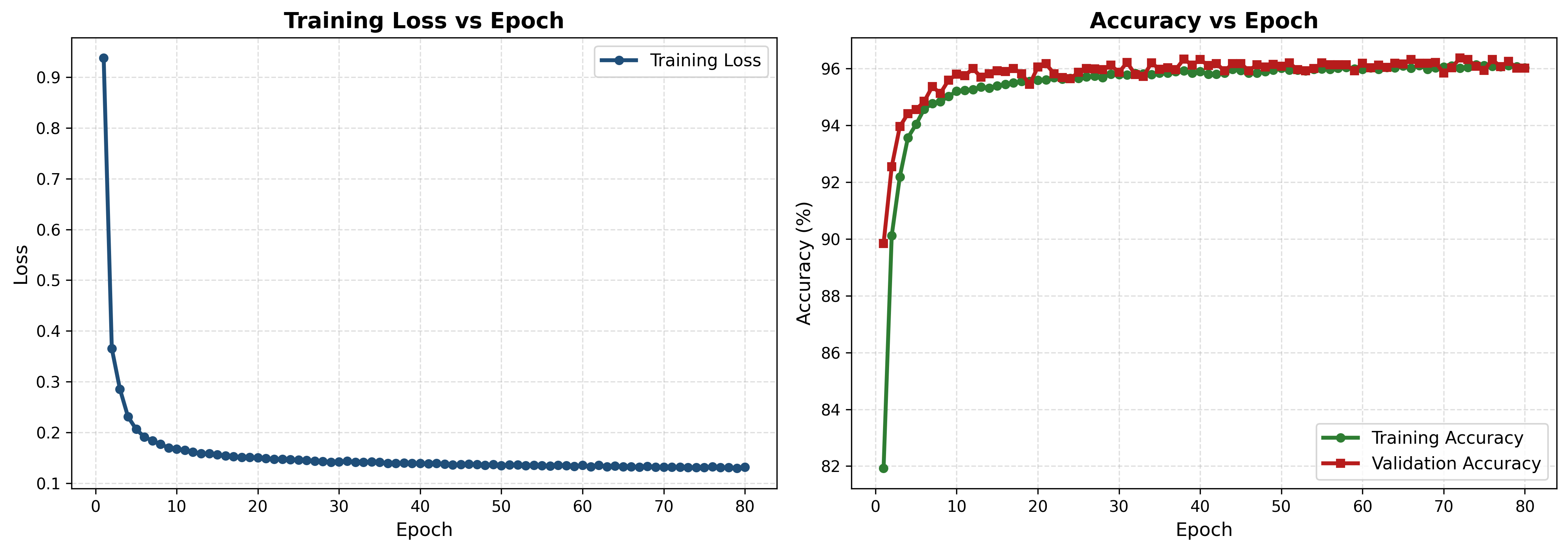}
        \caption{MNIST dataset.}
        \label{fig:fft_mnist}
    \end{subfigure}
    \caption{Training-loss and accuracy curves of the FD-CNN.}
    \label{fig:FFT_Training}
\end{figure*}

\begin{figure*}[t]
    \centering
    \begin{subfigure}[t]{0.48\textwidth}
        \centering
        \includegraphics[width=\linewidth]{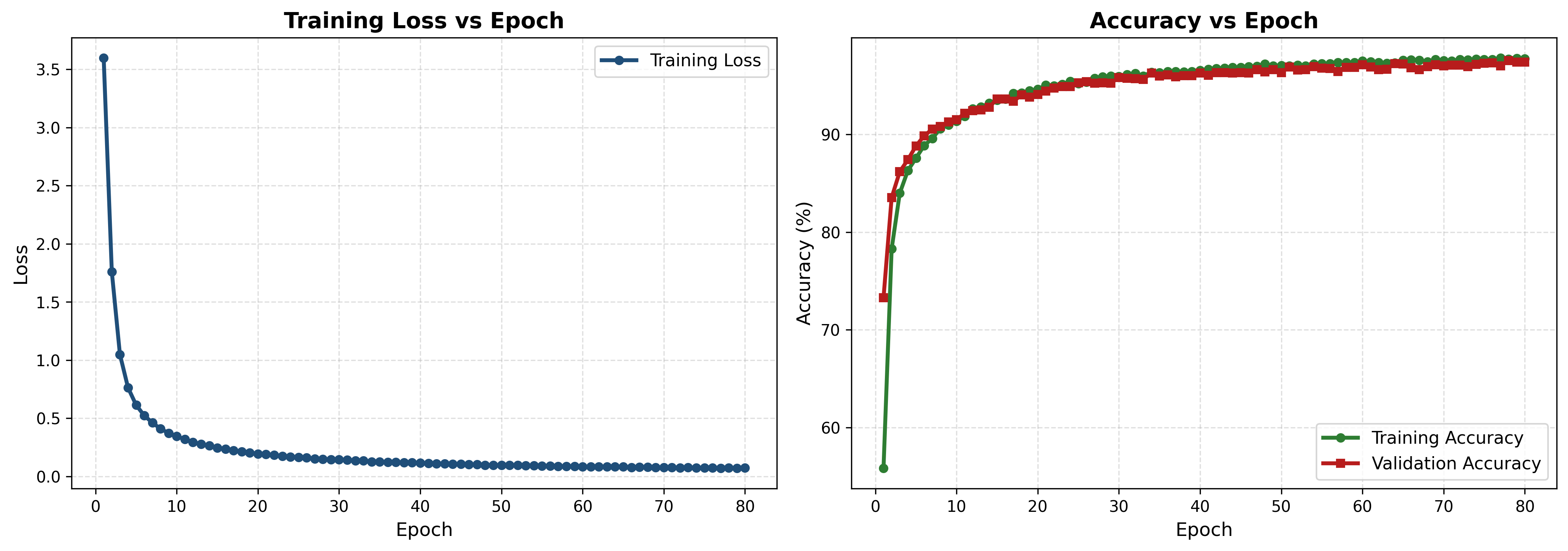}
        \caption{ASCII dataset.}
        \label{fig:dct_ascii}
    \end{subfigure}
    \hfill
    \begin{subfigure}[t]{0.48\textwidth}
        \centering
        \includegraphics[width=\linewidth]{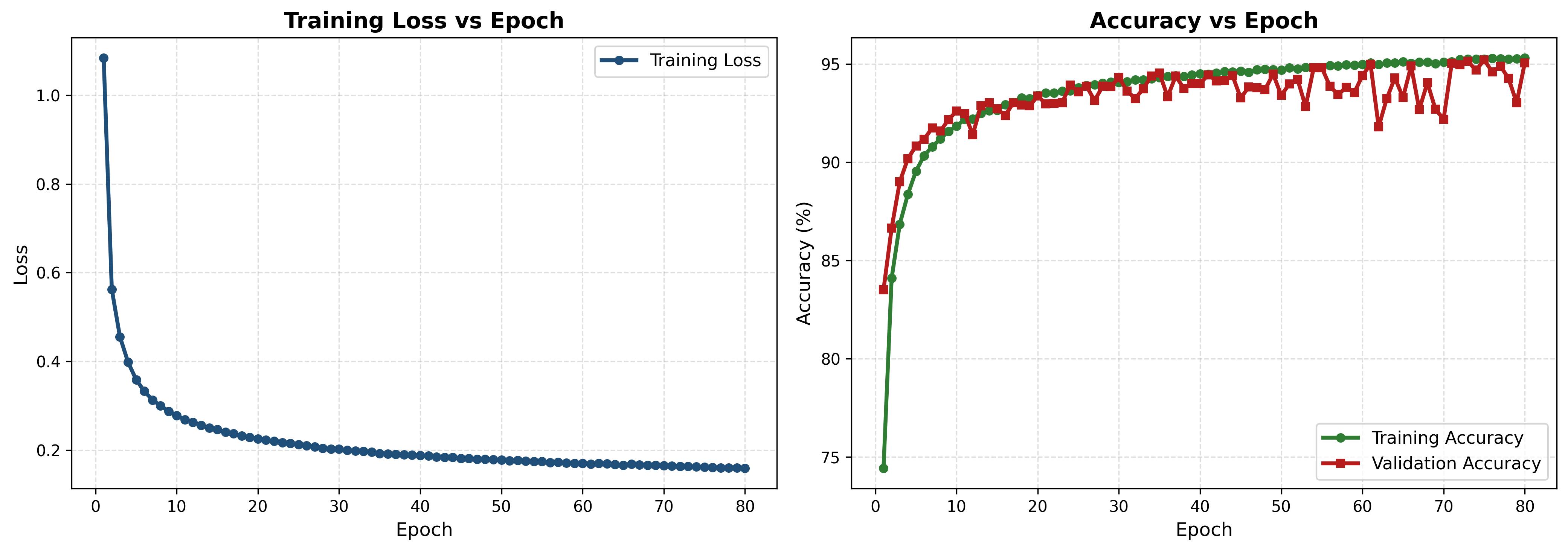}
        \caption{MNIST dataset.}
        \label{fig:dct_mnist}
    \end{subfigure}
    \caption{Training-loss and accuracy curves of the CD-CNN.}
    \label{fig:DCT_Training}
\end{figure*}

\subsubsection{Energy Efficiency}

Energy efficiency (EE) is defined as the number of inferences completed per joule:
\begin{equation}
    EE=
    \frac{\mathrm{Throughput}_{\mathrm{FPS}}}{P}
    \label{eq:energy_efficiency}
\end{equation}
where $P$ is the estimated total power in watts. The resulting unit is frames/J.

\section{Results}

The reference models are trained and evaluated using a workstation with a 12th-generation Intel Core i7 processor, 64 GB RAM, and an NVIDIA GeForce RTX 3060 GPU with 8 GB memory. The ASCII dataset is available at \texttt{https://github.com/\allowbreak VeCAD-P04-UTM/\allowbreak 94-Class-ASCII-Image-OCR-Data}.
 The CNN architectures are synthesized and implemented using AMD Xilinx Vitis HLS and Vivado 2025.2. The hardware performance is evaluated based on the post-implementation reports from AMD Xilinx Vivado 2025.2.

\subsection{Algorithmic Performance Evaluation}

The FD-CNN and CD-CNN training and loss curves are shown in Figures ~\ref{fig:FFT_Training} and \ref{fig:DCT_Training}. Table~\ref{tab:training_results} summarizes the classification results. On ASCII, the CD-CNN achieves a test accuracy of 97.39\%, compared with 97.78\% for the FD-CNN, a difference of 0.39 percentage points. On MNIST, the respective test accuracies are 95.05\% and 96.02\%, a difference of 0.97 percentage points. These results show that the CD-CNN maintains comparable classification accuracy while enabling real-valued DCT-domain processing.

\begin{table*}[t]
\centering
\caption{Training results of the evaluated SpCNN models on MNIST and ASCII.}
\label{tab:training_results}
\footnotesize
\begin{tabular}{llcccl}
\toprule
\textbf{Model} & \textbf{Dataset} & \textbf{Train Acc. (\%)} &
\textbf{Test Acc. (\%)} & \textbf{Loss} & \textbf{Training time}\\
\midrule
FD-CNN & ASCII & 97.82 & 97.78 & 0.054 & 1 h 12 min\\
CD-CNN & ASCII & 97.74 & 97.39 & 0.072 & 0 h 39 min\\
FD-CNN & MNIST & 96.03 & 96.02 & 0.132 & 3 h 14 min\\
CD-CNN & MNIST & 95.29 & 95.05 & 0.159 & 1 h 27 min\\
\bottomrule
\end{tabular}
\end{table*}

\subsection{Hardware Performance Evaluation}

Both architectures are implemented on the same FPGA platform under the stated implementation constraints. The CD-CNN uses real-valued DCT-domain processing, whereas the FD-CNN uses complex-valued FFT-domain processing. The following results compare their resource requirements, inference performance, power consumption, and energy efficiency.

\subsubsection{Hardware Resource Utilization}

The CD-CNN reduces normalized FPGA resource utilization from 25.82\% to 11.74\% for ASCII and from 17.35\% to 9.87\% for MNIST, corresponding to reductions of approximately 54.5\% and 43.1\%, respectively. These reductions are consistent with the lower storage and arithmetic requirements of real-valued spectral processing. In particular, the CD-CNN does not require separate imaginary-component processing, reducing associated datapath and intermediate-storage requirements. The magnitude of the reduction depends on the dataset and the resulting implementation configuration. Table~\ref{tab:RESOURCE} presents the FPGA resource utilization of the fully spectral FD-CNN and CD-CNN for both datasets.

\begin{table*}[t]
\centering
\caption{FPGA resource utilization across the SpCNN architectures.}
\label{tab:RESOURCE}
\footnotesize
\begin{tabular}{llccccc}
\toprule
\textbf{Model} & \textbf{Dataset} & \textbf{BRAM} &
\textbf{DSP} & \textbf{FF} & \textbf{LUT} & \textbf{Utilization (\%)}\\
\midrule
FD-CNN & ASCII & 441 & 227 & 25,976 & 53,100 & 25.82\\
CD-CNN & ASCII & 303 & 184 & 12,760 & 16,792 & \textbf{11.74}\\
FD-CNN & MNIST & 405 & 226 & 25,364 & 28,417 & 17.35\\
CD-CNN & MNIST & 163 & 183 & 12,723 & 16,697 & \textbf{9.87}\\
\bottomrule
\end{tabular}

\vspace{3pt}
\parbox{\textwidth}{\scriptsize
Utilization(\%): CLB-normalized resource utilization on XCKU5P FPGA. }
\end{table*}

\subsubsection{Computational Performance}

Table~\ref{tab:THROUGHPUT} compares inference performance under a 10-ns clock constraint. For ASCII, the CD-CNN reduces the RTL-simulation processing interval from 1,805,076 to 1,401,904 cycles and the reported latency from 14.71 to 11.48 ms. Throughput increases from 67.98 to 87.09 FPS. For MNIST, the processing interval decreases from 1,309,457 to 1,267,504 cycles, while latency decreases from 10.89 to 10.67 ms. Throughput increases from 91.82 to 93.72 FPS. The results show that the CD-CNN improves inference throughput on both datasets, with a larger improvement for ASCII.

\begin{table}[t]
\centering
\caption{Hardware performance of the evaluated SpCNN architectures.}
\label{tab:THROUGHPUT}
\scriptsize
\setlength{\tabcolsep}{4pt}
\begin{tabular}{lcc|cc}
\toprule
& \multicolumn{2}{c|}{\textbf{ASCII}} &
  \multicolumn{2}{c}{\textbf{MNIST}}\\
\textbf{Metric} & \textbf{FD-CNN} & \textbf{CD-CNN} &
\textbf{FD-CNN} & \textbf{CD-CNN}\\
\midrule
Clock constraint (ns) & 10 & 10 & 10 & 10\\
Interval (cycles) & 1,805,076 & 1,401,904 & 1,309,457 & 1,267,504\\
Latency (ms) & 14.71 & 11.48 & 10.89 & 10.67\\
$T_{\min}$ (ns) & 8.15 & 8.19 & 8.32 & 8.42\\
Throughput (FPS) & 67.98 & 87.09 & 91.82 & 93.72\\
$f_{\max}$ (MHz) & 123 & 122 & 120 & 119\\
\bottomrule
\end{tabular}
\end{table}

\begin{table}[t]
\centering
\caption{Power and energy metrics across the SpCNN architectures.}
\label{tab:POWER}
\scriptsize
\begin{tabular}{llccc}
\toprule
\textbf{Model} & \textbf{Dataset} & \textbf{P (W)} &
\textbf{Energy (mJ)} & \textbf{EE (frames/J)}\\
\midrule
FD-CNN & ASCII & 1.85 & 27.21 & 36.74\\
CD-CNN & ASCII & 0.86 & 9.87 & \textbf{101.27}\\
FD-CNN & MNIST & 1.23 & 13.39 & 74.65\\
CD-CNN & MNIST & 0.83 & 8.85 & \textbf{112.91}\\
\bottomrule
\end{tabular}
\vspace{3pt}
\parbox{\textwidth}{\scriptsize
P: Post-implementation total power with default switching activity.}

\end{table}

\subsubsection{Power and Energy Efficiency}

Table~\ref{tab:POWER} reports the power, per-inference energy, and energy efficiency of the two architectures. For ASCII, the CD-CNN improves energy efficiency by $2.76\times$, from 36.74 to 101.27 frames/J. For MNIST, it improves energy efficiency by $1.51\times$, from 74.65 to 112.91 frames/J. The improvements are associated with the combination of reduced power consumption and increased throughput. Real-valued operations remove the need for imaginary-component processing, while the actual power benefit also depends on resource mapping and switching activity.

\section{Comparison with Previous Works}

Table~\ref{tab:comp} compares the proposed CD-CNN implementations with prior FFT-domain and spatial CNN accelerators in terms of resource utilization, throughput, power consumption, and energy efficiency. Previous FD-CNN implementations in \cite{niu2019spec2} demonstrate high resource consumption due to complex real-imaginary data processing. While they achieve a frame rate of 148 on a Xilinx Virtex-7 FPGA using the 1000-class ImageNet dataset, existing literature lacks alternative CD-CNN hardware implementations for a direct baseline comparison, with available FD-CNN works \cite{zhang2017frequency, zeng2018framework, sun2018throughput,liu2022design, liu2020optimizing} employing Altera FPGAs.

Employing the same MNIST dataset and HLS methodologies, the spatial CNN implementations in \cite{cho2020implementation} and \cite{peng2025design} both yield high throughput. However, the architecture in \cite{peng2025design} demonstrates lower energy efficiency due to its elevated power consumption. Furthermore, these previous works report estimated resource and latency results derived directly from HLS, whereas this work provides post-implementation results validated through the Xilinx Vivado design suite and RTL simulation.

The fully spectral CD-CNN implemented on the GPU by Alshareef et al. \cite{alshareef2025real} achieves high throughput but comparatively low energy efficiency due to its power consumption. In contrast, the proposed CD-CNN implementation using the MNIST dataset improves energy efficiency by $12.96\times$ compared to the GPU implementation, demonstrating the potential of real-valued spectral processing for energy-efficient FPGA inference. Although the selected studies differ in datasets, target platforms, numerical precision, and evaluation methodologies, they represent the relevant implementations identified in the literature and are therefore included to provide a comparative assessment of the proposed architecture.

\begin{table*}[htpb]
\caption{Comparison with Similar Works}
\label{tab:comp}
\centering
\small
\setlength{\tabcolsep}{6pt}
\begin{tabular}{lcccccc}
\toprule
\textbf{Work} & HiPC'19 & ICEIC'20 & Array'25 & ICISCAE'25 & \textbf{Proposed} & \textbf{Proposed}\\
 & \cite{niu2019spec2} & \cite{cho2020implementation} & \cite{alshareef2025real} & \cite{peng2025design} &  &  \\
\midrule

\textbf{Model} & FD-CNN & CNN & CD-CNN & CNN & CD-CNN & CD-CNN\\
\textbf{Device} & Virtex-7 & XCZU9EG & GPU & KC270 & XCKU5P & XCKU5P\\

\textbf{Dataset} & ImageNet & MNIST & MNIST & MNIST & MNIST & ASCII \\

\textbf{Precision} & FX16 & FX20 & FT64 & - & FT32 & FT32 \\
\textbf{Design Entry} & RTL & HLS & Python & HLS & HLS & HLS \\

\hline
\textbf{DSP} & 3,200 & 143 & - & 46 & 183 & 184 \\
\textbf{LUT}(k) & 237 & 32.5  & - & 2.7 & 16.6 & 16.7 \\
\textbf{BRAM} & 1,244 & 95 & - & 18 & 163 & 303\\

\hline

\textbf{Throughput (FPS)} & 148 & 279.33 & 836.13 & 166 & \textbf{93.72} & \textbf{87.09} \\

\textbf{Power (W)} & - & - & 95.98 & 1.8 & \textbf{0.83} & \textbf{0.86}\\

\textbf{EE (frames/J)} & - & - & 8.71 & 92.22 & \textbf{112.91} & \textbf{101.27} \\

\bottomrule
\end{tabular}
\vspace{3pt}
\parbox{\textwidth}{\scriptsize
FT: Floating-Point Precision;
FX: Fixed-point Precision.}
\end{table*}

\section{Conclusion}

This work demonstrates the feasibility of an end-to-end, fully spectral CD-CNN accelerator on FPGA using HLS, targeting reduced hardware cost and improved energy efficiency for edge-AI inference. By combining real-valued DCT-domain processing with spectral convolution and pooling, the proposed architecture eliminates intermediate inverse transforms and reduces FPGA resource utilization by roughly $2\times$ compared with FD-CNN, for both the ASCII and MNIST datasets. Experimental results on MNIST and 94-class ASCII demonstrate energy efficiencies of 112.91 and 101.27 frames/J, achieving 1.51$\times$ and 2.76$\times$ improvements, respectively, while maintaining comparable classification accuracy. These findings establish DCT-domain processing as an alternative to complex-valued Fourier-domain architectures for resource-constrained FPGA deployment, providing a promising foundation for quantization and hardware optimizations.

\section*{Acknowledgements}

This research was supported by Universiti Teknologi Malaysia (UTM) under the Professional Development Research University Special Grant, number R.J130000.7113.07E51.

\bibliographystyle{elsarticle-num}
\bibliography{Reference}

@INPROCEEDINGS{he2020fft,
  author={He, Liulu and Xie, Xiaoru and Lin, Jun and Wang, Zhongfeng},
  booktitle={2020 IEEE Asia Pacific Conference on Circuits and Systems (APCCAS)}, 
  title={Efficient FPGA design for Convolutions in CNN based on FFT-pruning}, 
  year={2020},
  volume={},
  number={},
  pages={27-30},
  doi={10.1109/APCCAS50809.2020.9301653}}

@article{alshareef2025lightweight,
  title={Lightweight and high-precision spectral convolutional neural network model for custom 94-class ASCII character recognition},
  author={Alshareef, Ibrahim Y and Ab Rahman, Ab Al-Hadi and Khan, Nuzhat and Rizvi, Shahriyar Masud and Manzak, Ali and Rusli, Mohd Shahrizal and Mohammed, Mohammed Sultan and Hassan, Mohamed Khalafalla},
  journal={Neural Computing and Applications},
  pages={1--22},
  year={2025},
  publisher={Springer}
}

@article{liu2022design,
  title={Design of fully spectral cnns for efficient fpga-based acceleration},
  author={Liu, Shuanglong and Fan, Hongxiang and Luk, Wayne},
  journal={IEEE Transactions on Neural Networks and Learning Systems},
  volume={35},
  number={6},
  pages={8111--8123},
  year={2022},
  publisher={IEEE}
}

@article{rippel2015spectral,
  title={Spectral representations for convolutional neural networks},
  author={Rippel, Oren and Snoek, Jasper and Adams, Ryan P},
  journal={Advances in neural information processing systems},
  volume={28},
  year={2015}
}

@article{rizvi2023computation,
  title={Computation and memory optimized spectral domain convolutional neural network for throughput and energy-efficient inference},
  author={Rizvi, Shahriyar Masud and Rahman, Ab Al-Hadi Ab and Sheikh, Usman Ullah and Fuad, Kazi Ahmed Asif and Shehzad, Hafiz Muhammad Faisal},
  journal={Applied Intelligence},
  volume={53},
  number={4},
  pages={4499--4523},
  year={2023},
  publisher={Springer}
}

@article{rizvi2021low,
  title={A low-complexity complex-valued activation function for fast and accurate spectral domain convolutional neural network},
  author={Rizvi, Shahriyar Masud and Ab Rahman, Ab Al-Hadi and Khalil-Hani, Mohamed and Ayat, Sayed Omid},
  journal={Indonesian Journal of Electrical Engineering and Informatics (IJEEI)},
  volume={9},
  number={1},
  pages={173--184},
  year={2021}
}

@article{ayat2019spectral,
  title={Spectral-based convolutional neural network without multiple spatial-frequency domain switchings},
  author={Ayat, Sayed Omid and Khalil-Hani, Mohamed and Ab Rahman, Ab Al-Hadi and Abdellatef, Hamdan},
  journal={Neurocomputing},
  volume={364},
  pages={152--167},
  year={2019},
  publisher={Elsevier}
}

@inproceedings{xu2021dct,
  title={Dct-based fast spectral convolution for deep convolutional neural networks},
  author={Xu, Yuhao and Nakayama, Hideki},
  booktitle={2021 International Joint Conference on Neural Networks (IJCNN)},
  pages={1--8},
  year={2021},
  organization={IEEE}
}

@article{fulop2022end,
  title={End-to-end training of deep neural networks in the fourier domain},
  author={F{\"u}l{\"o}p, Andr{\'a}s and Horv{\'a}th, Andr{\'a}s},
  journal={Mathematics},
  volume={10},
  number={12},
  pages={2132},
  year={2022},
  publisher={MDPI}
}

@article{chen2021review,
  title={Review of image classification algorithms based on convolutional neural networks},
  author={Chen, Leiyu and Li, Shaobo and Bai, Qiang and Yang, Jing and Jiang, Sanlong and Miao, Yanming},
  journal={Remote Sensing},
  volume={13},
  number={22},
  pages={4712},
  year={2021},
  publisher={MDPI}
}

@inproceedings{mathieu2014fast,
  title={Fast training of convolutional networks through FFTs},
  author={Mathieu, Michael and Henaff, Mikael and LeCun, Yann},
  booktitle={International Conference on Learning Representations (ICLR)},
  year={2014}
}

@inproceedings{zhang2017frequency,
  title={Frequency domain acceleration of convolutional neural networks on CPU-FPGA shared memory system},
  author={Zhang, Chi and Prasanna, Viktor},
  booktitle={Proceedings of the 2017 ACM/SIGDA International Symposium on Field-Programmable Gate Arrays},
  pages={35--44},
  year={2017}
}

@inproceedings{zeng2018framework,
  title={A framework for generating high throughput CNN implementations on FPGAs},
  author={Zeng, Hailong and Chen, Ruizhe and Zhang, Chi and Prasanna, Viktor},
  booktitle={Proceedings of the 2018 ACM/SIGDA International Symposium on Field-Programmable Gate Arrays},
  pages={117--126},
  year={2018}
}

@article{abtahi2018accelerating,
  title={Accelerating convolutional neural network with FFT on embedded hardware},
  author={Abtahi, Tooraj and Kulkarni, Akshay and Mohsenin, Tinoosh},
  journal={IEEE Transactions on Very Large Scale Integration (VLSI) Systems},
  volume={26},
  number={9},
  pages={1737--1749},
  year={2018},
  publisher={IEEE}
}

@inproceedings{sun2018throughput,
  title={Throughput-optimized frequency domain CNN with fixed-point quantization on FPGA},
  author={Sun, Weiyi and Zeng, Hanqing and Yang, Yi-hua Edward and Prasanna, Viktor},
  booktitle={2018 International Conference on ReConFigurable Computing and FPGAs (ReConFig)},
  pages={1--8},
  year={2018},
  organization={IEEE}
}

@inproceedings{niu2019spec2,
  title={SPEC2: Spectral sparse CNN accelerator on FPGAs},
  author={Niu, Yue and Zeng, Hanqing and Srivastava, Ajitesh and Lakhotia, Kartik and Kannan, Rajgopal and Wang, Yanzhi and Prasanna, Viktor},
  booktitle={2019 IEEE 26th International Conference on High Performance Computing, Data, and Analytics (HiPC)},
  pages={195--204},
  year={2019},
  organization={IEEE}
}

@inproceedings{liu2020optimizing,
  title={Optimizing fully spectral convolutional neural networks on FPGA},
  author={Liu, Shuanglong and Luk, Wayne},
  booktitle={2020 International Conference on Field-Programmable Technology (ICFPT)},
  pages={39--47},
  year={2020},
  organization={IEEE}
}

@inproceedings{zeng2017fast,
  title={Fast generation of high throughput customized deep learning accelerators on FPGAs},
  author={Zeng, Hanqing and Zhang, Chi and Prasanna, Viktor},
  booktitle={2017 International Conference on ReConFigurable Computing and FPGAs (ReConFig)},
  pages={1--8},
  year={2017},
  organization={IEEE}
}

@article{lahti2025high,
  title={High-level Synthesis for FPGAs-A Hardware Engineer’s Perspective},
  author={Lahti, Sakari and H{\"a}m{\"a}l{\"a}inen, Timo D},
  journal={IEEE Access},
  year={2025},
  publisher={IEEE}
}

@article{alshareef2025real,
  title={A real-valued DCT-based spectral CNN architecture for efficient edge deep learning},
  author={Alshareef, Ibrahim Yousef and Ab Rahman, Ab Al-Hadi and Hassan, Mohamed Khalafalla and Rusli, Mohd Shahrizal and Khan, Nuzhat and Manzak, Ali and Marsono, Muhammad Nadzir},
  journal={Array},
  pages={100594},
  year={2025},
  publisher={Elsevier}
}

@article{alshareef2026relu,
  title={ReLU-DCT and LASP: Real-valued frequency-domain activations and adaptive pooling for DCT-Based CNNs},
  author={Alshareef, Ibrahim Yousef and Ab Rahman, Ab Al-Hadi and Shaameri, Ahmad Zuri and Khan, Nuzhat and Rusli, Mohd Shahrizal and Hassan, Mohamed Khalafalla and Manzak, Ali and Marsono, Muhammad Nadzir},
  journal={Knowledge-Based Systems},
  pages={116018},
  year={2026},
  publisher={Elsevier}
}

@article{alshareef2026end,
  title={End-to-end discrete cosine transform integration in spectral convolutional neural networks for resource-efficient deep learning},
  author={Alshareef, Ibrahim Yousef and Ab Rahman, Ab Al-Hadi and Khan, Nuzhat and Alqaraghuli, Hasan},
  journal={Applied Soft Computing},
  pages={114599},
  year={2026},
  publisher={Elsevier}
}

@article{alshareef2025optimizing,
  title={Optimizing spectral activation functions for spectral convolutional neural networks: balancing efficiency and fidelity},
  author={Alshareef, Ibrahim and Ab Rahman, Ab Al-Hadi and Shahrizal, Mohd and Alqaraghuli, Hasan and Khan, Nuzhat},
  journal={ELEKTRIKA-Journal of Electrical Engineering},
  volume={24},
  number={2},
  pages={176--183},
  year={2025}
}

@article{TRETS2025HLS,
author = {Gottschaldt, Paul and Podlubne, Ariel and G{\"o}hringer, Diana},
title = {A Taxonomy of the High-Level Synthesis Ecosystem for Heterogeneous FPGA Systems},
year = {2025},
issue_date = {December 2025},
publisher = {Association for Computing Machinery},
address = {New York, NY, USA},
volume = {18},
number = {4},
issn = {1936-7406},
url = {https://doi.org/10.1145/3764664},
doi = {10.1145/3764664},
journal = {ACM Trans. Reconfigurable Technol. Syst.},
month = oct,
articleno = {44},
numpages = {33}
}

@manual{amd_ug574,
  author       = {{Advanced Micro Devices, Inc.}},
  title        = {UltraScale Architecture Configurable Logic Block User Guide},
  organization = {Advanced Micro Devices, Inc.},
  number       = {UG574 (v1.6)},
  year         = {2025},
  month        = {jan},
  url          = {https://docs.amd.com/r/en-US/ug574-ultrascale-clb/CLB-Resources},
  note         = {Accessed: Oct. 5, 2026}
}

@inproceedings{cho2020implementation,
  title={Implementation of data-optimized FPGA-based accelerator for convolutional neural network},
  author={Cho, Mannhee and Kim, Youngmin},
  booktitle={2020 International Conference on Electronics, Information, and Communication (ICEIC)},
  pages={1--2},
  year={2020},
  organization={IEEE}
}

@inproceedings{peng2025design,
  title={Design and Research on LeNet5 Convolutional Neural Network Accelerator Based on HLS},
  author={Peng, Chengyi},
  booktitle={2025 IEEE 8th International Conference on Information Systems and Computer Aided Education (ICISCAE)},
  pages={726--731},
  year={2025},
  organization={IEEE}
}

@article{ju2022convolutional,
  title={Convolutional neural networks with discrete cosine transform features},
  author={Ju, Sanghyeon and Lee, Youngjoo and Lee, Sunggu},
  journal={IEEE Transactions on Computers},
  volume={71},
  number={12},
  pages={3389--3395},
  year={2022},
  publisher={IEEE}
}

@inproceedings{ito2022convolution,
  title={Convolution using discrete cosine transforms for improving performance of convolutional neural networks},
  author={Ito, Izumi},
  booktitle={2022 30th European Signal Processing Conference (EUSIPCO)},
  pages={1462--1466},
  year={2022},
  organization={IEEE}
}

\end{document}